\documentclass[submission,copyright,creativecommons]{eptcs}
\providecommand{\event}{AREA 2020} % Name of the event you are submitting to

\usepackage[acronym, nomain, nonumberlist]{glossaries}
\usepackage{amsmath}
\usepackage{placeins}

\usepackage{graphicx}
\usepackage{subfig}
\usepackage{siunitx}

\usepackage{listings}

\usepackage{breakurl}             % Not needed if you use pdflatex only.
\usepackage{underscore}           % Only needed if you use pdflatex.
\usepackage{subfiles}           % Best loaded last in the preamble

\graphicspath{{images/}{../images/}}
\newacronym{ABM}{ABM}{Agent Based Modelling}
\newacronym{AI}{AI}{Artificial Intelligence}
\newacronym{AOP}{AOP}{Agent Oriented Programming}
\newacronym{API}{API}{Application Programming Interface}
\newacronym{ASL}{ASL}{AgentSpeak(L) Language}
\newacronym{BDI}{BDI}{Belief-Desire-Intention}
\newacronym{DEVS}{DEVS}{Discrete Event System Specification}
\newacronym{GNSS}{GNSS}{Global Navigation Satellite System}
\newacronym{GPIO}{GPIO}{General-Purpose Input/Output}
\newacronym{GPS}{GPS}{Global Positioning System}
\newacronym{HLA}{HLA}{High Level Architecture}
\newacronym{IR}{IR}{Infrared}
\newacronym{JADE}{JADE}{Java Agent Development Framework}
\newacronym{MAS}{MAS}{Multi Agent System}
\newacronym{OODA}{OODA}{Observe Orient Decide Act}
\newacronym{PROFETA}{PROFETA}{Python RObotic Framework for dEsigning sTrAtegies}
\newacronym{QR}{QR code}{Quick Response code}
\newacronym{ROS}{ROS}{Robot Operating System}
\newacronym{SAVI}{SAVI}{Simulated Autonomous Vehicle Infrastructure}
\newacronym{SOIFRA}{SOIFRA}{Service-Oriented Interoperable Framework for Robot Autonomy}
\newacronym{TCP}{TCP}{Transmission Control Protocol}
\newacronym{UAV}{UAV}{Unmanned Aerial Vehicle}
\title{Toward Campus Mail Delivery Using BDI}

\author{Chidiebere Onyedinma
\institute{University of Ottawa\\ Ontario, Canada}
\email{conye066@uottawa.ca}
\and
Patrick Gavigan \qquad\qquad Babak Esfandiari  
\institute{Carleton Univeristy\\
Ontario, Canada}
\email{\quad patrickgavigan@sce.carleton.ca \quad\qquad babak@sce.carleton.ca }
}

\def\titlerunning{Toward Campus Mail Delivery Using BDI}
\def\authorrunning{Onyedinma, Gavigan \& Esfandiari}  % Title, author details

\begin{document}
    \maketitle
    % !TeX spellcheck = en_GB
\begin{abstract}
Autonomous systems developed with the \gls{BDI} architecture are usually mostly implemented in simulated environments. In this project we sought to build a \gls{BDI} agent for use in the real world for campus mail delivery in the tunnel system at Carleton University. Ideally, the robot should receive a delivery order via a mobile application, pick up the mail at a station, navigate the tunnels to the destination station, and notify the recipient.

We linked the \gls{ROS} with a \gls{BDI} reasoning system to achieve a subset of the required use cases. \gls{ROS} handles the low-level sensing and actuation, while the BDI reasoning system handles the high-level reasoning and decision making. Sensory data is orchestrated and sent from ROS to the reasoning system as perceptions. These perceptions are then deliberated upon, and an action string is sent back to ROS for interpretation and driving of the necessary actuator for the action to be performed.

%When a user orders a robot via a mobile application for mail delivery, the robot should meet the user at the specified pick up post point, receive the mail, navigate through the tunnel, and deliver it at the specified delivery location. All this should happen in perception-action cycles through ROS and BDI.

In this paper we present our current implementation, which closes the loop on the hardware-software integration, and implements a subset of the use cases required for the full system.
\end{abstract}

\glsresetall
    % !TeX spellcheck = en_GB
\section{Introduction} \label{sec:Introduction}
Autonomous systems are designed in such a manner that they can intelligently react to ever-changing environments and operational conditions. Given such flexibility, they can accept goals and set a path to achieve these goals in a self-responsible manner while displaying some form of intelligence. An autonomous agent can be defined as a system that senses the environment and acts on it, over time, in pursuit of its own agenda and to affect what it senses in the future \cite{JasonBook}.

The \gls{BDI} framework is meant for developing such autonomous agents, in that it defines how to select, execute and monitor the execution of user-defined plans (Intentions) in the context of current perceptions and internal knowledge of the agent (Beliefs) in order to satisfy the long-term goals of the agent (Desires). But so far, very few applications of \gls{BDI} have been observed outside of simulated or virtual environments. In this paper, we describe how we built our autonomous robot that uses \gls{BDI} (and specifically, the Jason implementation of the \gls{BDI} \gls{ASL}) and \gls{ROS} to eventually deliver interoffice mail in the Carleton University campus tunnels. The Carleton tunnel system allows people to go from one campus building to another without having to face Ottawa's harsh winters, and makes for a fairly controlled environment that our iRobot platform can navigate. However, being underground also means that access to GPS is not possible, and internet access is limited to certain areas. In this context, ultimately the robot will have to know where it is and where to deliver mail, but there are also some sub-goals, like obstacle avoidance and battery recharge, which it might have to achieve in other to get to its main goal.

In the remainder of the paper, we first provide some background on \gls{BDI}, \gls{ROS}, and related work on known implementations of agent-based robots; we then describe our overall hardware and software architecture; next, we  describe in more detail the hardware and software implementation; finally, we describe the few test cases that we have implemented so far using the completed robot. 
    % !TeX spellcheck = en_GB

\section{Background} \label{sec:Background}
We provide background on \gls{BDI} and the \gls{ASL} in section \ref{sec:BeliefDesireIntentionArchitecture}. We then introduce \gls{ROS} in section \ref{sec:RobotOperatingSystem} followed by a discussion of various related work in section \ref{sec:RelatedWork}.

\subsection{Belief-Desire-Intention Architecture} \label{sec:BeliefDesireIntentionArchitecture}

The principles that underpin \gls{BDI} originated in the 1980s cognitive science theory as a means of modelling agency in humans \cite{bratman1987intention}. Since that time, this model has been applied in the development of software agents as well as the field of \gls{MAS}. An example of a popular implementation of applying \gls{BDI} to agent reasoning is Jason \cite{JasonBook} \cite{JasonWeb}. In Jason, an agent's initial belief base, goals and plans are specified using \gls{ASL}. This language is also used for communication between agents.

In \gls{BDI} systems, a software agent performs reasoning based upon internally held \emph{beliefs}, stored in a belief base, about itself and the task environment. The agent also has objectives, or \emph{desires}, that are provided to it, as well as a plan base, which contains various means for achieving goals depending on the agent's context. The agent's reasoning cycle consists of first perceiving the task environment, and receiving any messages. From this information, the agent can then decide on a course of action suitable to the context provided by those perceptions, the agent's own beliefs, messages received, and desires. Once this course of action has been selected, we can say that the agent has set an \emph{intention} for itself. These plans can include updating the belief base, sending messages to other agents, and taking some action in the task environment. As the agent continues to repeat its reasoning cycle, it can reassess the applicability of its intentions as it perceives the environment, and drop intentions that are no longer applicable \cite{JasonBook} \cite{JasonWeb}.

%\subsection{AgentSpeak} \label{sec:AgentSpeak}
Agents developed for \gls{BDI} systems using Jason are programmed using a language called \gls{ASL}. This is a logic based programming language that bears similarities to Prolog. The syntax provides a means for specifying initial beliefs for the agent to have, rules that can be applied for reasoning as well as plans that can be executed.

\lstset{
  caption=Example AgentSpeak program. \label{lst:ExampleAgentSpeakProgram},
  basicstyle=\scriptsize, frame=tb,
  xleftmargin=.05\textwidth, xrightmargin=.05\textwidth
}
\begin{lstlisting}
destination(post1).
atDestination :- destination(DESTINATION) & postPoint(DESTINATION,_).
+!goToLocation : atDestination <- drive(stop).
\end{lstlisting}

The \gls{ASL} syntax is illustrated in listing \ref{lst:ExampleAgentSpeakProgram}. First, we have a simple belief in the form of a predicate, in this case it is the knowledge that the destination (of our robot) is \texttt{post1}. Next, we have a rule following the format of \emph{implication}, where the $:-$ operator holds the meaning of the implication arrow ($\implies$) in reverse. In this case, we have a rule defining \texttt{atDestination}, which is implied when \texttt{DESTINATION} unifies to the current location specified by the \texttt{postPoint(CURENT,PREVIOUS)} perception. Finally, we have a plan, which follows the form: 
\begin{verbatim}
    triggeringEvent : context <- body. 
\end{verbatim}
A \emph{triggering event} can be the addition or deletion of a belief, achievement goal, or a test goal. Achievement goals begin with an exclamation mark (!), test goals begin with a question mark (?) and beliefs do not begin with punctuation. Triggers that are based on the addition or deletion of a belief or goal begin with a positive (+) or negative (-) sign respectively. An achievement goal is used for providing the agent with an objective with respect to the state of the environment whereas a test goal is generally used for querying the state of the environment. The \emph{context} is a set of conditions that must be satisfied in order for the plan to be applicable based on the state of the agent's belief base. This is a logical sentence that can use both beliefs as well as previously defined rules. The \emph{body} includes the instructions for the agent to follow for executing the plan. The plan body can include the addition or deletion of beliefs and/or goals as well as actions for the agent to perform \cite{JasonBook} . In our example, we have a plan for the achievement goal of \texttt{!goToLocation}. This plan sends the action \text{drive(stop)} when the \texttt{atDestination} context rule is satisfied.

% \lstset{
%   caption=AgentSpeak syntax. \label{lst:AgentSpeakSyntax},
%   basicstyle=\scriptsize, frame=tb,
%   xleftmargin=.05\textwidth, xrightmargin=.05\textwidth
% }
% \begin{lstlisting}
% A.
% B :- A.
% triggeringEvent : context <- body.
% \end{lstlisting}

\subsection{Robot Operating System}\label{sec:RobotOperatingSystem}
\gls{ROS} is a package for developing software for robotic applications \cite{ROSWeb}. This system operates using a tuple-space architecture. Various software nodes publish and subscribe to various \emph{topics} using socket-based communications. This is managed using a central master node which has the role of brokering peer-to-peer connections between nodes that publish and subscribe to the same topics. This allows developers to focus on the implementation of individual nodes and enables flexibility to use one of many available nodes that are compatible with \gls{ROS}. For example, various hardware component developers have made \gls{ROS} nodes available, allowing systems developers to use those modules without concern as to how those nodes are implemented in detail. \gls{ROS} also provides functionality for recording runtime data, which can be used for diagnostics.

\subsection{Related Work}\label{sec:RelatedWork}
Although there are many examples of research on software agents and the use of \gls{BDI}, this review of related work focuses on the application of \gls{BDI} to robotics where the development was targeted toward real-world applications. We will also discuss work that sought to use \gls{BDI} agents with \gls{ROS}.

The Australian military conducted research into the use of \gls{BDI} for controlling a fixed-wing \gls{UAV} called a Codarra Avatar. As part of this project, they developed both the ``Automated Wingman'', a graphical programming environment where pilots could provide mission-specific programming for a \gls{UAV}, as well as a \gls{BDI}-based flight controller for the \gls{UAV} itself. The intent of this research was to enable pilots, who may not have programming skills, to provide mission parameters in a way more natural to them using the military's \gls{OODA} loop. The authors proposed that the \gls{OODA} loop could be approximated using \gls{BDI}. Successful flight tests were performed using these systems in the mid 2000s, although it is unclear if any follow-on research was conducted \cite{TheAutomatedWingmanUsingJackIntelligentAgentsForUnmannedAutonomousVehicles} \cite{ExperiencesWithTheDesignAndImplementationOfAnAgentBasedAutonomousUAVController}.

A more recent example of \gls{BDI} being used for controlling a drone was provided by Menegol \cite{EvaluationOfMultiAgentCoordinationOnEmbeddedSystems} \cite{msmenegolUAVExperiments}. Their implementation uses the JaCaMo framework \cite{JaCaMoWeb}, which includes Jason. A video of their \gls{UAV} flying is available online \cite{JasonAgentDrone}. This work is currently being extended to use the \gls{ROS} as the core of the architecture \cite{JasonROSGitHub} \cite{RezendersMASUAV}. Their approach is to build a linkage between \gls{ROS} and Jason, where Jason agents can run actions by passing messages to modules in \gls{ROS} and receive perceptions by receiving messages from other modules. The perceptions and actions are defined using manifest files that specify the properties and parameters of the messages. This is similar to other efforts to link \gls{ROS} to Jason, such as rason \cite{rason}, JaCaROS \cite{JaCaROSGithub}, and JROS \cite{JROS}, although it is unclear if these efforts are related to this project.

Taking another approach using Python, the \gls{PROFETA} library implements \gls{BDI} and the AgentSpeak language designed for use with autonomous robots \cite{APythonFrameworkForProgrammingAutonomousRobotsUsingADeclarativeApproach}. They are interested in determining if \gls{AOP} can be implemented with Python for simpler robotic implementations. In their paper, the authors used the Eurobot challenge as well as a simulated warehouse logistics robot scenario as case studies. In the Eurobot challenge, the robot must sort objects in the environment while also working in the presence of other, uncooperative, robots \cite{EurobotWeb}. 

%The \gls{SOIFRA} is an agent implementation for robotics intended to be interoperable for both air and ground vehicles \cite{InteroperableMultiAgentFrameworkForUnmannedAerialGroundVehiclesTowardsRobotAutonomy}. This framework uses a layered approach for providing agency to robotic platforms using a \gls{MAS} where different agents, implemented using \gls{JADE}, are responsible for different aspects of the robot's behaviour.

The ARGO project \cite{ARGOAnExtendedJasonArchitectureThatFacilitatesEmbeddedRoboticAgentsProgramming} has interfaced Jason agents with Arduino using a library called Javino \cite{ARoboticAgentPlatformForEmbeddingSoftwareAgentsUsingRaspberryPiAndArduinoBoards}. Javino is a Java library for controlling Arduino computers from Java programs that was specifically designed with the intention of using it to control a robot using Jason programs. That being said, the authors of the ARGO paper claim to not be tied to specific hardware or a specific \gls{AOP} language, such as AgentSpeak \cite{ARGOAnExtendedJasonArchitectureThatFacilitatesEmbeddedRoboticAgentsProgramming} \cite{ARoboticAgentPlatformForEmbeddingSoftwareAgentsUsingRaspberryPiAndArduinoBoards}.

Alzetta and Giorgini contributed work toward a real-time \gls{BDI} system connected to \gls{ROS} 2 \cite{TowardsARealTimeBDIModelForROS2} \cite{ROS2BDI}. Their implementation uses a custom built \gls{BDI} engine, implemented in C++, which supports soft real-time constraints. The agent's \emph{desires} are encoded with soft real-time deadlines for when they need to be achieved. The plans in the agent's plan library include the execution time for that plan. The agent reasoning system can then reason about the priority of desires, time constraints and execution time when performing plan selection.

Finally, the authors' own related work includes the \gls{SAVI} project, which aimed to develop an architecture for simulating autonomous agents implemented using a Jason \gls{BDI} \cite{SAVIPaper} \cite{SAVIWeb}. Among its key features is the decoupling of the agent reasoning cycle from the simulation time cycle, enabling the simulated agents to run in their own time. The agent's perceptions and actions passed between the simulated agent body running in a separate thread and decoupled from the agent's reasoning cycle. Although this system was targeted toward a simulated environment, the design was intended to be useful for application to robotic agents, not only simulated agents.
    % !TeX spellcheck = en_GB
\section{Motivation} \label{sec:Motivation}

%\subsection{Why BDI} \label{sec:WhyBdi}
Our use of \gls{BDI} for this work is two fold. First, \gls{BDI} provides a good goal oriented agent architecture that is resilient to plan failure and changes to context. It also supports the notion of shorter-term and longer term plans that can be organized so as not to conflict with each other. \gls{BDI} is also inherently a social agent system, providing useful utilities for message passing and communication in \gls{MAS}, which relates to future work for this project.  Granted, \gls{BDI} may not necessarily be the \emph{perfect} ad hoc solution for agent based robotics, but there's just not enough literature to demonstrate the appropriateness of \gls{BDI} (or lack thereof) in robotics.

There are alternative agent architectures to \gls{BDI} that are available, for example the subsumption architecture \cite{SubsumptionPaper} \cite{AnIntroductionToMultiAgentSystems}. Although it is likely possible that the robotic behaviours implemented in this paper likely could have implemented the same robot using subsumption, our longer term goals for this project would likely make the use of other architectures more difficult. \gls{BDI} is inherently social and goal directed whereas in the case of the subsumption architecture, the agent behaviour emerges from the various layers built into the agent \cite{IntegratedCognitiveArchitecturesASsurvey}.

%\subsection{How we differ to related work}
Our goal is to use an established \gls{BDI} system, namely Jason in an ecosystem for various robotic platforms (\gls{ROS}) and enable to use of agent systems to solve real-world problems using robotics, taking advantage of \gls{ROS}' ecosystem of publishers and subscribers. As mentioned in section \ref{sec:Background}, while there are some projects that have sought to control real-world robotics using \gls{BDI} reasoning systems, there are a limited number of works in this area. Those that are available differ from our work in a number of ways. For example, in the case of the Codarra avatar agent, although it is very interesting, it does not seem to be openly available. Other work, such as \gls{PROFETA}, uses a Python based \gls{BDI}, as opposed to the more commonly used Jason. Our work is more similar in motivation to the efforts to link Jason and \gls{ROS} mentioned in section \ref{sec:Background}, although our implementation of the connection between \gls{ROS} and Jason is quite different. In our case, the \gls{BDI} reasoning system is built as a stand alone program with rosjava using Jason as a library, without the use of an external middleware.

    % !TeX spellcheck = en_GB
\section{Architecture} \label{sec:Architecture}
This section outlines the architecture of the mail delivery robot. The robot is intended to function on an on-demand basis, where a mail-sending user would summon the robot to collect mail, similar to how users request rides using ride-sharing apps. The robot would then autonomously navigate to a nearby mail collection and delivery location to collect the item from the user. Once the mail has been collected, the robot would then autonomously navigate to the mail delivery location and alert the receiver that there is mail for them to receive. The receiver would then meet the robot at another mail collection and delivery location. For the purposes of this early stage prototype, the mail delivery locations, and any other points of interest are indicated using a \gls{QR}, and the robot paths are marked using a line for the robot to follow. Removing the need for instrumenting the environment will be discussed in the future work, in section \ref{sec:FutureWork}.

First, in section \ref{sec:Environment}, we examine the task environment that the robot will operate in. We then discuss the hardware configuration in section \ref{sec:Hardware}. The software architecture is discussed in section \ref{sec:Software}.

\subsection{Environment} \label{sec:Environment}
The task environment for the robot is the tunnel system that connects the buildings of Carleton University. This provides our robot with an indoor space with no weather to deal with and smooth floors to drive on, and also connects to many buildings on campus. Although these are attractive features of the tunnel system, there are some drawbacks. First, the tunnels do not have consistent wireless internet coverage, although there are locations where there is reliable network access. The tunnels also have lower lighting levels than typical office environments, providing a potential challenge to the design. Finally, in the tunnel there is no access to \gls{GNSS} signals, such as \gls{GPS}, meaning that the robot will need to determine its location another way.

\subsection{Hardware} \label{sec:Hardware}
\begin{figure}[!htbp]
    \centering
    \includegraphics[width=8cm]{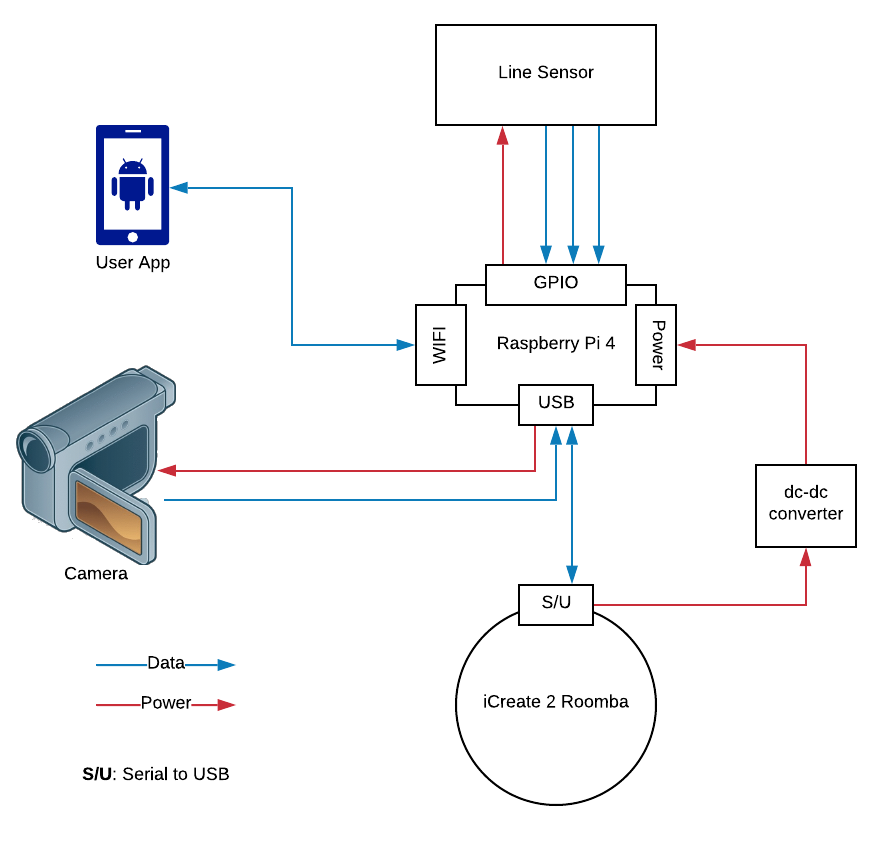}
    \caption{Mail delivery robot hardware.}
    \label{fig:MailDeliveryRobotHardware}
\end{figure}

The hardware configuration of the mail delivery robot is shown in figure \ref{fig:MailDeliveryRobotHardware}. The mail delivery robot is primarily implemented using an iRobot Create2, which is the development version of the Roomba vacuum cleaning robot, without the vacuum-cleaning components. This robot can be controlled using a command protocol over a serial interface \cite{RoombaWeb}, and can also be used to provide power to other connected devices. A Raspberry Pi 4 computer was attached to the robot and connected via a serial cable, and powered from the robot's battery using a power adapter. Also connected via a serial connection are a camera and a line sensor used for detecting a line on the floor of the tunnels.

\subsection{Software} \label{sec:Software}
\begin{figure}[!htbp]
    \centering
    \includegraphics[width=8.5cm]{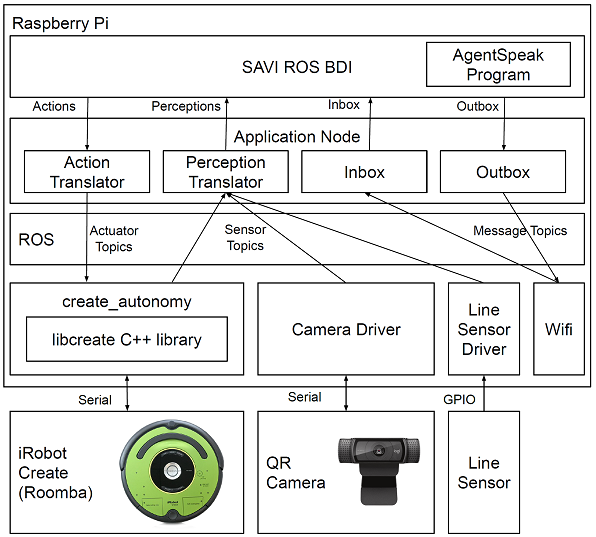}
    \caption[Mail delivery robot software architecture]{Mail delivery robot software architecture (robot image credit \cite{RoombaWeb}, camera image credit: \cite{webcamWeb}).}
    \label{fig:MailDeliveryRobotSoftwareArchitecture}
\end{figure}

\begin{figure}[!htbp]
    \centering
    \includegraphics[width=11cm]{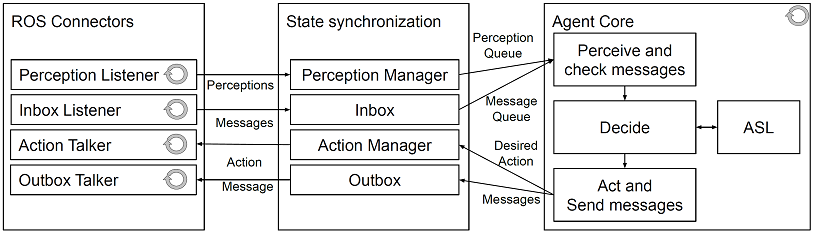}
    \caption{SAVI ROS BDI internal architecture.}
    \label{fig:SaviRosBdiInternalArchitecture}
\end{figure}

The control software is implemented using a set of modules connected via \gls{ROS}, as shown in figure \ref{fig:MailDeliveryRobotSoftwareArchitecture}. The reasoning system for this robot, inspired by the \gls{SAVI} project \cite{SAVIPaper,SAVIWeb}, decouples the reasoning cycle from the interface to the sensors and actuators using a state synchronization module. The internal reasoning system for this project, called \emph{\gls{SAVI} \gls{ROS} \gls{BDI}}, and shown in figure \ref{fig:SaviRosBdiInternalArchitecture}, is inspired by the original \gls{SAVI} configuration. Implemented in Java, using the rosJava package \cite{rosjavaWeb} and the Jason \gls{BDI} engine \cite{JasonBook} \cite {JasonWeb}, this module connects to \gls{ROS} directly, subscribing to perceptions and inbox messages and publishing actions and outbox messages as required. Again, the state synchronization module is important as perceptions and messages can arrive at any time, decoupled from the reasoning cycle of the agent. This is set up in three main components: The \gls{ROS} connectors, the state synchronization module, and the agent core. The \gls{ROS} connectors are responsible for subscribing to either perceptions or inbox messages, or publishing actions or outbox messages, each in their own thread of execution. These are connected to the state synchronization module, which manages queues or messages in and out of the agent as well as perceptions and actions in and out of the agent.  The agent core, which runs the agent reasoning cycle in a separate thread of execution, checks for perceptions and inbox messages at the beginning of the reasoning cycle. Then, the agent decides on an appropriate course of action and then updates the agent state with new outbox messages and actions which need to be executed. The agent behaviour is defined by an \gls{ASL} file which is parsed by the reasoning system at start-up, making this module fully platform agnostic: there are no assumptions about the underlying hardware, capabilities, or mission of the agent in the implementation of this system. This agent reasoning system is available at \cite{saviRosBdiWeb}.

The Create2 robot platform can use the \texttt{create\_autonomy} package available in \gls{ROS}, which connects to an underlying C++ library called \texttt{libcreate} to \gls{ROS}, publishing the data from various sensors as \gls{ROS} topics and subscribing to topics related to the various commands available to the robot \cite{CreateAutonomyWeb}. Also connected in this way are drivers for the \gls{QR} camera and photodiode line sensor, which each publish their data as \gls{ROS} topics. Also connected to \gls{ROS} is \gls{SAVI} \gls{ROS} \gls{BDI}, described above. Lastly, as required by \gls{SAVI} \gls{ROS} \gls{BDI}, is the application node translator, which reformats sensor data as AgentSpeak perceptions and conversely translates action commands in AgentSpeak to the relevant topics being subscribed to by the \texttt{create\_autonomy} package. Lastly, an AgentSpeak program is provided to the reasoning system, which defines the behaviour of the agent. The implementation of the perception and action translators, the drivers for the \gls{QR} camera and the line sensor, and the \gls{ASL} program are available at \cite{saviRoombaWeb}.

    \section{Implementation} \label{sec:Implementation}
\begin{figure}[!htbp]
	\centering
	\subfloat[Side view of the robot. \label{fig:SideViewOfTheRobot}]{{\includegraphics[width=7cm]{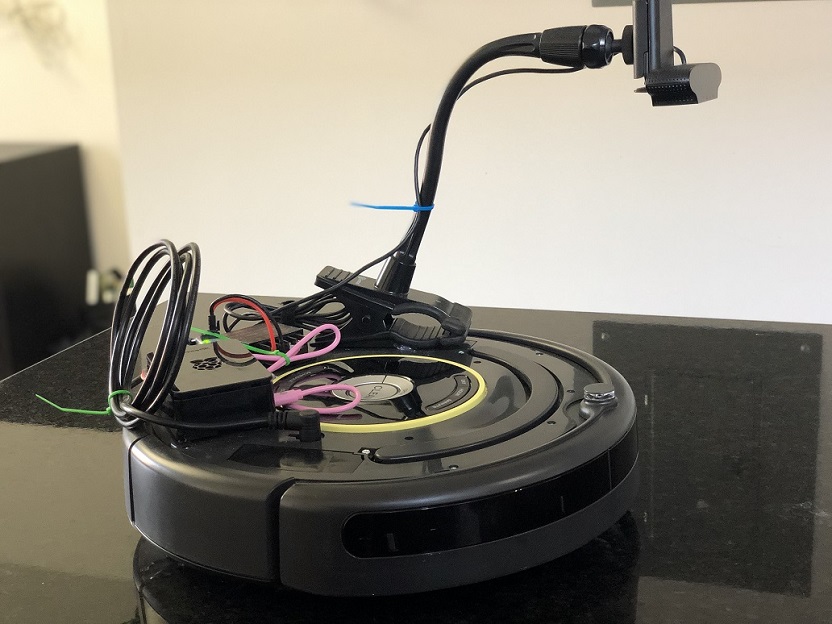} }}
	\qquad
	\subfloat[Top view of the robot. \label{fig:TopViewOfTheRobot}]{{\includegraphics[width=7cm]{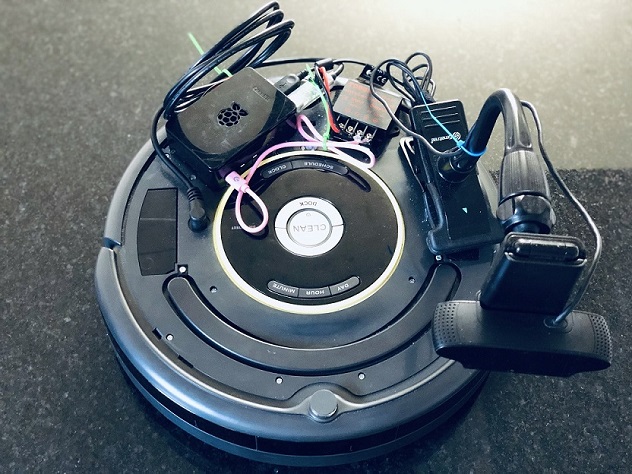} }}
	\caption{Assembled robot prototype.}
	\label{fig:AssembledPrototype}
\end{figure}

This section discusses the implementation of the various aspects of the system, shown in figure \ref{fig:AssembledPrototype}. The source code for this project can be found on GitHub \cite{saviRoombaWeb,saviRoombaAppWeb}. First, we show how we powered the on-board computer using the robot's power system, in section \ref{sec:SystemPower}. Next we discuss the means of maneuvering the robot using line sensing, including the implementation of the line sensor and associated driver in section \ref{sec:ManeuveringWithLineSensing}. The robot uses a system based on \gls{QR} for determining its position, given the lack of other methods such as \gls{GNSS}. This is discussed in section \ref{sec:LocationSensingWithQRCodes}. The user interface, implemented as an Android app, is discussed in section \ref{sec:UserInterface}. The action translator, which handles the implementation of the robot's actuators is explained in section \ref{sec:ActionTranslator}. Finally, the details of the implementation of the agent behaviour, in \gls{ASL} are provided in section \ref{sec:AgentBehaviour}.

\subsection{System Power} \label{sec:SystemPower}
\begin{figure}[!htbp]%
	\centering
	\subfloat[RS232 pinout. \label{fig:rs232Pinout}]{{\includegraphics[width=3cm]{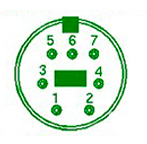} }} %
	\qquad
	\subfloat[USB pinout. \label{fig:USBpinout}]{{\includegraphics[width=6cm]{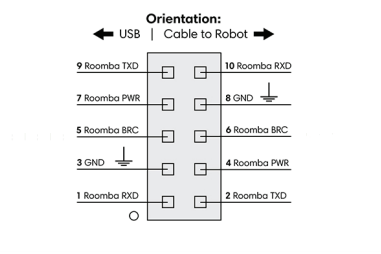} }}%
	\caption{Serial connection pinout \cite{CreatePinout}.}%
	\label{fig:SerialConnectionPinout}%
\end{figure}

In order to power the robot's computer (a Raspberry Pi 4) without having it tethered to a socket in a wall, we utilized the iRobot Create 2's power system. This was possible as the serial connection between the computer and the robot also provides access to the robot's internal rechargeable battery. Conveniently, the serial cable used to connect the Create 2 to the Raspberry Pi exposes the robot's power bus through its RS232 pinout, as seen in figure \ref{fig:SerialConnectionPinout} and table \ref{table:pinout}.

\begin{table}[!htbp]
    \centering
    \caption{Create 2 external serial port RS232 connector pinout.}
    \label{table:pinout}
    \begin{tabular}{|c|c|c|}
    \hline
    Pin & Name & Description \\ [0.5ex] 
    \hline
    1 & Vpower & battery + (unregulated) \SIrange{16}{20}{V} \\ 
    2 & Vpower & battery + (unregulated) \SIrange{16}{20}{V} \\
    3 & RXD & \SIrange{0}{5}{V} Serial input to robot \\
    4 & TXD & \SIrange{0}{5}{V} Serial output from robot \\
    5 & BRC & Baud Rate Change \\ 
    6 & GND & battery ground \\
    7 & GND & battery ground \\ [1ex] 
    \hline
    \end{tabular}
\end{table}

Although this pinout provides access to the robot's power supply, it must be converted from 16-20 V to the regulated \SI{5}{V} required by the Raspberry Pi computer via its USB-C connector or its \gls{GPIO} pin. To get a stable \SI{5}{V} for our Raspberry Pi, we used a Tobsun 15W DC to DC power converter by feeding power to its input (\SI{12}{V}/\SI{24}{V} positive and negative) terminal from pin 4 and pin 3 of the Serial to USB header described in figure \ref{fig:USBpinout} respectively. We connected the exposed wires of an improvised USB type C cable to the converter and then we plugged in the cable to the Raspberry Pi; when the RS232 end of the Serial-to-USB cable is plugged into the Create2 robot, the entire system is powered successfully.

% \begin{figure}[!htbp]
%     \centering
%     \includegraphics[width=8cm]{DC to DC conversion}
%     \caption{DC to DC Power conversion for Raspberry Pi.}
%     \label{fig:DCToDCPowerConversionRaspberry}
% \end{figure}

% \begin{figure}[!htbp]
%     \centering
%     \includegraphics[width=7cm]{usb}
%     \caption{USB internal wiring.}
%     \label{fig:USBInternalWiring}
% \end{figure}

With the robot and computer successfully powered by the robot's power supply, it is necessary for the reasoning system to have awareness of the battery charge state, so that it can report to a charging station if necessary. The \texttt{create_autonomy} package regularly publishes a \gls{ROS} topic called \texttt{battery/chargeratio} which indicates the percentage of charge left on the battery based on its capacity. The perception translator node, implemented in Python, subscribes to this topic and publishes a \texttt{batteryOK} or \texttt{batteryLow} string to the \texttt{perceptions} \gls{ROS} topic. A \texttt{batteryOK} is published when the charge left in the battery is above 25\% and a \texttt{batteryLow} otherwise. If the robot receives a \texttt{batteryLow} perception, it drops all other plans and looks for the closest docking station in order to recharge itself, explained in more detail in section \ref{sec:AgentBehaviour}.

\subsection{Maneuvering with Line Sensing}\label{sec:ManeuveringWithLineSensing}
\begin{figure}[!htbp]
    \centering
    \includegraphics[width=0.45\textwidth]{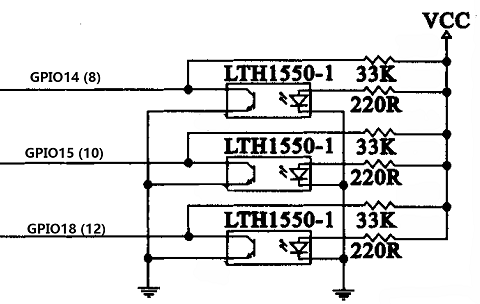}
    \caption{Line sensor circuit.}   
    \label{fig:LineSensorSchematic}
\end{figure}

As our robot operates in an indoor environment without the support of \gls{GNSS} systems for navigation, a simple means of moving through the tunnels and navigation was required. As an initial implementation, a line sensor was used for the robot to follow lines on the tunnel floor. This sensor is implemented using three Photo-interrupter LTH 1550-01 diodes, shown in figure \ref{fig:LineSensorSchematic}. Each sensor detects if the line is on the left, center or right of the robot's center. Two resistors were used per Photo-interrupter, a \SI{220}{\ohm} and a \SI{33}{\kilo\ohm}. The \SI{220}{\ohm} was used as a limiting resistor for the LED within the sensor and the \SI{33}{\kilo\ohm} as a voltage divider to enable us to measure the voltage across the resistor when light falls on the photo-transistor.

The sensors were connected to three different \gls{GPIO} pins on the Raspberry Pi. The right sensor is connected to GPIO14 (pin8), the center sensor to GPIO15 (pin10) and the left sensor to GPIO18 (pin12). The sensor is powered from the Raspberry Pi; the VCC pins are connected together and then to the 5v pin of the Raspberry Pi, while the ground (GND) pins are connected together and then to the ground (GND) pin of the Raspberry Pi. When light falls on each of these sensors, their GPIO pins are set to HIGH, and when the sensors are covered or faced with a non-reflective material or has no light falling on them, their GPIO pins are set to LOW. 

\begin{figure}[!htbp]
	\centering
	\subfloat[Robot base layout. \label{fig:RoombaBaseLayout}]{{\includegraphics[width=7cm]{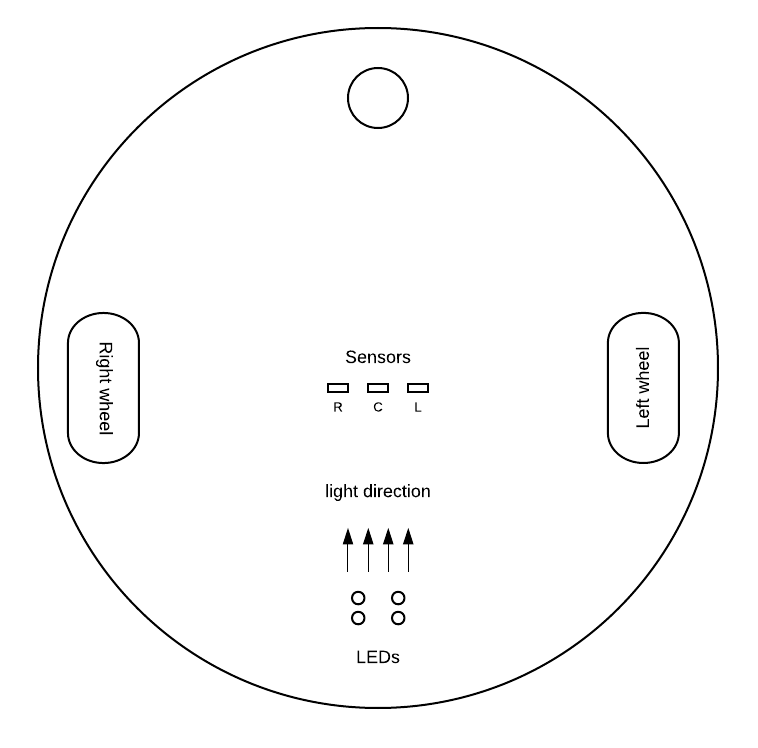} }}
	\qquad
	\subfloat[Under the robot. \label{fig:UnderTheRobot}]{{\includegraphics[width=7cm]{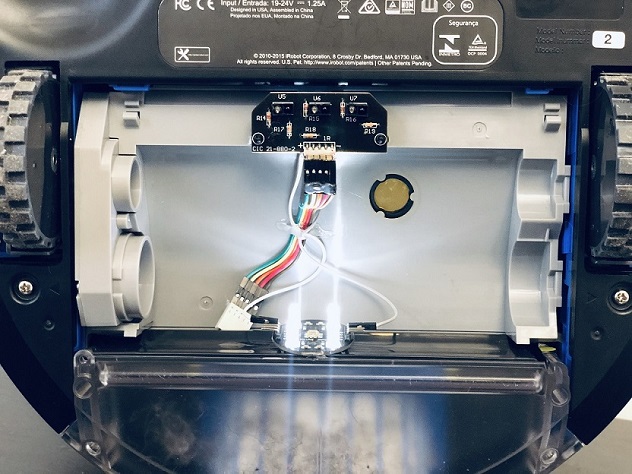} }}
	\caption{Layout of the underside of the robot.}
	\label{fig:LayoutOfTheUndersideOfTheRobot}
\end{figure}

The navigation track was designed using a reflective black tape, so that when it is faced by any of the sensors, the respective \gls{GPIO} pin is set to HIGH, and then we know if the line is on the right, center or left depending on the pin that was set to HIGH or LOW. The sensors are mounted under the center of the Create2, in line with the right and left wheels, as seen in figure \ref{fig:RoombaBaseLayout}. An image of the underside of the robot itself is provided in figure \ref{fig:UnderTheRobot}. This is to ensure more navigation accuracy, because if the sensors are mounted in front, or behind the wheels, the line would be detected before or after the robot needs to make a navigation decision. For example, if the sensors are mounted in front of the wheels, and while the robot is in motion (following the line) the line changes direction; the change in direction is detected first by the sensors making the robot to turn and change its direction before it needs to, thereby making it go out of track.

The robot has a broad surface area and when on the floor, has little or no light underneath it. Since our sensors are mounted under the robot, they cannot function effectively because they need a certain amount of light in order to detect the line. A light source under the robot using four LEDs was added. These LEDs were mounted perpendicular to the sensors with their light directed at the sensor. With this in place, when the robot is on the floor, the light bounces off any reflective object or material placed on the floor, and is absorbed by non-reflective materials or objects.

%\subsubsection{Line Track Design}\label{sec:LineTrackDesign}
The line tracks are created using tapes. To ensure enough contrast between the line to follow and the floor regardless of the environment, we had to create a track with two different types of tape, reflective and non-reflective, with the non-reflective tape in the center. This type of line track would work irrespective of location flooring. The robot is kept on the track, with the center sensor on the non-reflective tape, so when any of the sensors is faced with the non-reflective tape, we know the line track is in that direction.

\subsubsection{Sensing and Publishing Line Location}\label{sec:SensingAndPublishingLineLocation}
With the line sensor hardware implemented, we needed to consider how the signals would be sent to the \gls{BDI} reasoning system. This was implemented in a software driver implemented in Python (referred to as the line sensor driver in figure \ref{fig:MailDeliveryRobotSoftwareArchitecture}). This driver consists of a \gls{ROS} node which runs in a \SI{10}{Hz} loop, implemented using \gls{ROS}'s \texttt{rospy.Rate()} and \texttt{rospy.sleep()} functions, and interfaces with the hardware via the Raspberry Pi's \gls{GPIO} library. The software and monitors if the signals from the \gls{GPIO} pins are HIGH or LOW, indicating if the diodes of the line sensor are detecting the line under them. 

The \gls{ROS} node publishes strings to the \texttt{perceptions} topic. These include \texttt{line(center)}, meaning that the line was detected in the center of the sensor, \texttt{line(left)} and \texttt{line(right)} which means that the line was detected by the left or right sensor and possible also the center sensor. The perception \texttt{line(across)} is used to indicate that the line was detected by all of the sensors and the \texttt{line(lost)} indicates that the line is not visible to any of the sensors. These perceptions were then received by the \gls{BDI} reasoner and interpreted as part of the agent reasoning cycle, discussed in section \ref{sec:AgentBehaviour}.

\subsection{Location sensing with QR Codes}\label{sec:LocationSensingWithQRCodes}
As the tunnel system in which the robot is expected to operate has no access to external navigation systems, such as \gls{GNSS}, it was necessary for the robot to have another means of identifying its location. This was accomplished by posting \gls{QR} codes along the path of the robot but without obstructing the line track that the robot would be following. The camera used for scanning the codes was also positioned on the left side of the robot and ten inches from the floor because of its focal length; this was to enable the camera to capture the code properly. For two-directional travel, a QR code was placed on either side of the line, enabling the robot to detect a code that could also inform the robot of the direction of travel.% This is also shown in figure \ref{fig:TapeTracks}.

%\begin{figure}[!htbp]
%    \centering
%    \includegraphics[width=6cm]{double-qr}
%    \caption{QR placement.}
%    \label{fig:DoubleQRPlacement}
%\end{figure}

The \gls{QR} is scanned using a \gls{ROS} node which is responsible for managing the camera, called the camera driver in figure \ref{fig:MailDeliveryRobotSoftwareArchitecture}. Implemented in Python, the camera driver scans for \gls{QR} codes at \SI{2}{Hz} rate. When a  code is detected, the location number included in the image is published to the \texttt{perceptions} \gls{ROS} topic and saved in a location history in the camera driver. The format of this perception is: \texttt{postPoint(C, P)} where C is the current scanned location number, and P is the previously scanned location number. This predicate is received by the \gls{BDI} reasoning system and processed using the AgentSpeak rules discussed in more detail in section \ref{sec:AgentBehaviour}.

\subsection{User Interface}\label{sec:UserInterface}
The user interface currently consists of a rudimentary mobile application for the Android platform, and a Python script that resides on the robot, responsible for relaying messages from the user to the \texttt{perceptions} topic. Upon start up, a \gls{TCP} connection is established between these two applications. After a successful connection to the robot's communication node, the sender can then select the pickup post point and the destination post point. This is received by the node, and the destination is published to the perception's \gls{ROS} topic as a predicate string \texttt{dest(D)} where D is the post location number of the destination. The user interface is available online \cite{saviRoombaAppWeb}.

% \begin{figure}[!htbp]%
% 	\centering
% 	\subfloat[App interface, connect to robot. \label{fig:ConnectRobot}]{{\includegraphics[width=5cm]{connect-robot} }}%
% 	\qquad
% 	\subfloat[App interface, send robot. \label{fig:SendRobot}]{{\includegraphics[width=5cm]{post-point-select} }}%
% 	\caption{User interface.}%
% 	\label{fig:UserInterface}%
% \end{figure}

\subsection{Action translator} \label{sec:ActionTranslator}
When the robot reasoning system requests that an action be performed by the robot, the action is published to the \texttt{actions} \gls{ROS} topic. These messages are interpreted by the \texttt{action translator}, a Python script which subscribes to the \texttt{actions} topic and then publishes messages to the appropriate topics for the \texttt{create_autonomy} node to control the lower level hardware of the robot. The action messages that are currently supported include actions for driving the robot forward, turning left and right, and stopping the robot. These commands are: \texttt{drive(forward)}, \texttt{turn(left)}, \texttt{turn(right)}, \texttt{drive(stop)}. We also support actions for docking and undocking the robot from the charging station using the internal programming of the robot: \texttt{dock\_bot} and \texttt{undock\_bot}.

%\begin{table}[!htbp]
%\centering
%\caption{Action message definitions.}
%\label{table:ActionMessageDefinitions}
%\begin{tabular}{|l|l|}
%\hline
%\multicolumn{1}{|c|}{Action message} & \multicolumn{1}{c|}{Intended meaning}      \\ \hline
%\texttt{drive(forward)}                                                                  & Drive robot forward                        \\ \hline
%\texttt{turn(left)}                                                                     & Turn robot to the left                     \\ \hline
%\texttt{turn(right)}                                                                    & Turn robot to the right                    \\ \hline
%\texttt{drive(stop)}                                                                     & Stop the robot                             \\ \hline
%\texttt{dock\_bot}                                                                       & Dock the robot with the charging station   \\ \hline
%\texttt{undock\_bot}                                                                     & Undock the robot from the charging station \\ \hline
%\end{tabular}
%\end{table}

\subsection{Agent behaviour} \label{sec:AgentBehaviour}
The reasoning system receives inputs via perceptions and actuates via actions based upon the results of its reasoning cycle. The agent behaviour is defined for the Jason \gls{BDI} reasoner in \gls{ASL}. The \gls{ASL} program uses a hierarchy of behavioural goals, each of which have supporting plans providing the agent with a means of achieving the goals in a given context. 

%The context is supported by a set of rules that are used to define the map in which the robot needs to operate, explained in section \ref{sec:MapDefinitionRulesAndBeliefs}. At the highest level, we have the goal of \texttt{!deliverMail}. The plans associated with achieving this goal are explained in detail in section \ref{sec:MailDeliveryPlans}. These plans invoke the \texttt{!goToLocation} goal for handling robot navigation. The plans for achieving this goal are defined in section \ref{sec:NavigationPlans}. In a similar fashion, the navigation plans delegate to the \texttt{!followPath} goal. The plans for path following are explained in section \ref{sec:PathFollowingPlans}. There is also the goal of docking the robot to recharge the battery. The plans for how to dock the robot are in section \ref{sec:DockingPlans}.

\begin{figure}[!htbp]
    \centering
    \includegraphics[width=5.5cm]{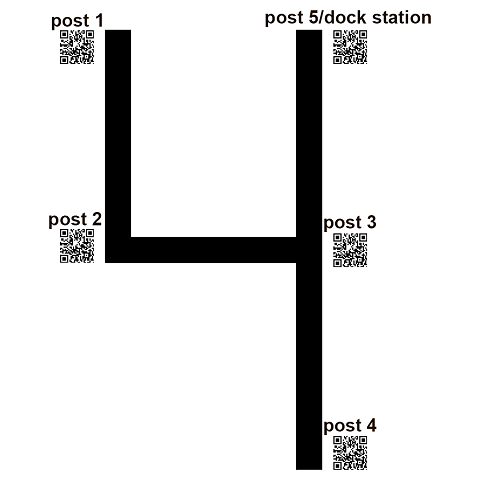}
    \caption{Map of the test environment.}
    \label{fig:MapOfTheTestEnvironment}
\end{figure}

%\subsubsection{Map definition rules and beliefs} \label{sec:MapDefinitionRulesAndBeliefs}
The testing environment map is shown in figure \ref{fig:MapOfTheTestEnvironment}. The plans for the navigation goals of the robot use rules defined for \texttt{atDestination}, \texttt{DestinationBehind}, \texttt{DestinationAhead}, \texttt{DestinationLeft}, and \texttt{DestinationRight}. In turn these rules depend on internally held beliefs within the robot which define the location of the mail sender, mail receiver and the docking station. These beliefs are structured in the form of predicates defined as \texttt{senderLocation(post1)}, \texttt{receiverLocation(post4)}, and \texttt{dockStation(post5)}, where the terms in brackets are locations on the map. These beliefs can be hard coded in the \gls{ASL} file or be provided to the agent via communication messages.

Examples of the rules used for this map are provided in listing \ref{lst:ExampleRulesForTheMap}. First we see the \texttt{atDestination} rule, where the agent should have the belief that the currently detected post point, provided by a perception, matches the \texttt{destination()} belief. In this case, the term \texttt{DESTINATION} must unify. The second term in the \texttt{postPoint()} is an underscore (\_) as this term defines the previously detected post point, which is not relevant to this rule. Next we have an example of a rule for \texttt{DestinationRight} for the scenario where the robot's destination is \texttt{post4}, \texttt{post2} was the previously detected location and is currently detecting \texttt{post3}. These and all other rules (which will be automatically generated based on the map configuration in our next iteration) can be found at \cite{saviRoombaWeb}.

\lstset{
  caption=Example rules for the map.\label{lst:ExampleRulesForTheMap},
  basicstyle=\scriptsize, frame=tb,
  xleftmargin=.05\textwidth, xrightmargin=.05\textwidth
}
\begin{lstlisting}
atDestination :- destination(DESTINATION) & postPoint(DESTINATION,_).
DestinationRight :- destination(DESTINATION) & postPoint(CURRENT,PAST) &
CURRENT = post3 & PAST = post2 & DESTINATION = post4.
\end{lstlisting}

%\subsubsection{Mail delivery plans} \label{sec:MailDeliveryPlans}
The highest level goal for this robot is \texttt{!deliverMail}. Listing \ref{lst:MailDeliveryPlans} shows some examples plans that are triggered by this goal. First, we see a plan for a scenario where the agent does not have \texttt{haveMail} in the knowledge base. Here, the context check unifies the location of the sender and verifies if the agent is already at the sender's location. The battery charge is also checked to make sure that there is sufficient power to continue. If this context is satisfied, the agent adds a predicate to set the destination as the sender's location. The agent then adopts the goal of \texttt{!goToLocation} and also readopts the goal of \texttt{!deliverMail}, as the goal of delivering the mail has not yet been achieved, also making this a recursive plan. The second plan that can trigger on this goal addresses the situation where the robot has arrived at the sender's location. In that situation, the robot adds the belief of \texttt{haveMail} to the knowledge base and also removes any previous destination. Again, this is recursive as the mail has not yet been delivered. Two additional, and very similar plans, are required for setting the destination as the receiver for the cases where the mail has been collected. Those plans are available at \cite{saviRoombaWeb}. Next we see the scenario where the battery has an insufficient charge. In this situation, the destination is set to the docking station.

\vspace*{-0.3cm}
\lstset{
  caption=Example mail delivery plans.\label{lst:MailDeliveryPlans},
  basicstyle=\scriptsize, frame=tb,
  xleftmargin=.05\textwidth, xrightmargin=.05\textwidth
}
\begin{lstlisting}
+!deliverMail : ((not haveMail) & senderLocation(SENDER) &
receiverLocation(RECEIVER) & not postPoint(SENDER,_) & batteryOK) 
<- +destination(SENDER); !goToLocation; !deliverMail.

+!deliverMail : ((not haveMail) & senderLocation(SENDER) & 
receiverLocation(RECEIVER) & postPoint(SENDER,_) & batteryOK) 
<- +haveMail; -destination(_); !deliverMail.

+!deliverMail : (batteryLow & dockStation(DOCK)) <- -destination(_); 
+destination(DOCK); !goToLocation; !deliverMail.
\end{lstlisting}

%\subsubsection{Navigation plans} \label{sec:NavigationPlans}
Listing \ref{lst:NavigationPlans} provides the plans for navigating the robot to the destination in order to achieve the goal of \texttt{!goToLocation}, which is invoked by the plans that achieve the goal of \texttt{!deliverMail}. These plans are fairly simple thanks to the rules previously defined for their context definitions. The plans use the actions of \texttt{drive(stop)} to stop if the robot is at the destination, and \texttt{turn()} to turn in the direction of the destination. Otherwise, the plans invoke the goal of \texttt{!followPath} to follow the line path, and are all recursive if the agent is not yet at the destination.

\lstset{
  caption=Navigation plans.\label{lst:NavigationPlans},
  basicstyle=\scriptsize, frame=tb,
  xleftmargin=.05\textwidth, xrightmargin=.05\textwidth
}
\begin{lstlisting}
+!goToLocation : destinationAhead <- !followPath.
+!goToLocation : atDestination <- drive(stop).
+!goToLocation : destinationLeft <- turn(left); !followPath.
+!goToLocation : destinationRight <- turn(right); !followPath.
\end{lstlisting}

%\subsubsection{Path following plans} \label{sec:PathFollowingPlans}
The plans for achieving the goal of \texttt{!followPath} are shown in listing \ref{lst:PathFollowingPlans}. The first plan commands the robot to drive forward when the line is detected in the center of the line sensor. Next we have a plan that stops the robot if the line is not visible. Lastly, there is a plan that turns the robot in the direction that the line is detected, using unification.

\vspace*{-0.3cm}
\lstset{
  caption=Path following plans.\label{lst:PathFollowingPlans},
  basicstyle=\scriptsize, frame=tb,
  xleftmargin=.05\textwidth, xrightmargin=.05\textwidth
}
\begin{lstlisting}
+!followPath : line(center) <- drive(forward); !followPath.
+!followPath : line(lost) <- drive(stop).
+!followPath : line(DIRECTION) <- drive(DIRECTION); !followPath.
\end{lstlisting}

%\subsubsection{Docking plans} \label{sec:DockingPlans}
In the situation where the robot needs to dock to charge the battery, the agent adopts the goal of \texttt{!dock}. The plans for achieving this goal are provided in listing \ref{lst:DockingPlans}. Here, we have two plans, the first for going to the dock location, if we are not already there, and the second for initiating the \texttt{dock_bot} action.

\lstset{
  caption=Docking plans.\label{lst:DockingPlans},
  basicstyle=\scriptsize, frame=tb,
  xleftmargin=.05\textwidth, xrightmargin=.05\textwidth
}
\begin{lstlisting}
+!dock : dockStation(DOCK) & not postPoint(DOCK,_) <- !goToLocation; !dock.
+!dock : dockStation(DOCK) & postPoint(DOCK,_) <- drive(stop); dock_bot.
\end{lstlisting}
    % !TeX spellcheck = en_GB
\section{Evaluation} \label{sec:Evaluation}
%\FloatBarrier

We evaluated our prototype robot in two ways: by observing it driving through a test course and by measuring the publication period of perceptions and actions to \gls{ROS} topics. Our interest was to assess if the \gls{BDI} reasoning system was running sufficiently fast in order to keep up with the perceptions being generated.
\vspace*{-0.3cm}
\begin{figure}[h]
    \centering
    \subfloat[Perception period \label{fig:PerceptionPeriod}]{{\includegraphics[width=7cm]{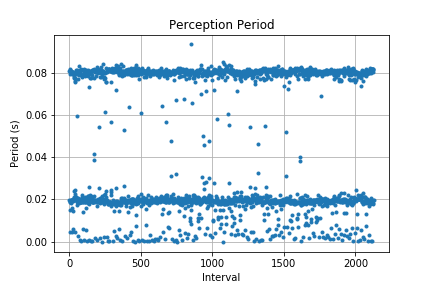}}}%
    \qquad
    \subfloat[Action period \label{fig:ActionPeriod}]{{\includegraphics[width=7cm]{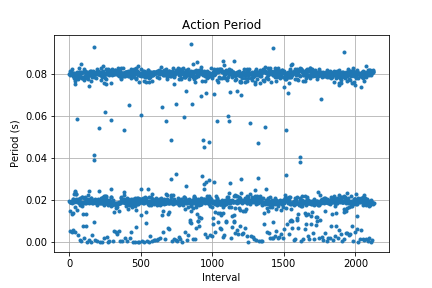}}}%
    \caption{Perception and action message periods.}%
    \label{fig:PerceptionAndActionMessagePeriods}%
\end{figure}
\vspace*{-0.3cm}
%A video of the robot operating in the test environment is available \cite{saviRoombaVideo}. In the video we see the robot executing a number of mail delivery tasks as well as interrupt the mail delivery so that it can charge the battery before continuing.

Figure \ref{fig:PerceptionAndActionMessagePeriods} shows the period at which perceptions and actions were published to their associated topics in \gls{ROS} as the robot was executing the mail delivery task. As various perceptions are published at different rates, we can see a strong tendency for there to be messages approximately every \SI{0.02}{s} and every \SI{0.08}{s}. As the reasoning system for \gls{BDI} generates an action every time a perception is received, we conversely see actions being published at approximately the same rate. This tells us that the \gls{BDI} reasoning system was able to keep up with the perceptions generated by the various sensors. Further investigation will include profiling our system to determine where it spends the most time.

    % !TeX spellcheck = en_GB
\section{Conclusion}\label{sec:Conclusion}
In this paper, we presented the work to date on the development of a robotic agent for performing autonomous mail delivery in a campus environment. We conclude with a description of our key accomplishments and a view toward our future work.

\subsection{Key Accomplishments}\label{sec:KeyAccomplishments}
We demonstrated the feasibility of using \gls{BDI} in an embedded system. We accomplished this using the \gls{SAVI} \gls{ROS} \gls{BDI} framework, linking Jason's \gls{BDI} reasoning system to \gls{ROS}. We also implemented our initial robot behaviours in \gls{BDI}, navigating through the environment using line-following and \gls{QR} while also monitoring the battery state, seeking a charging station as needed.

We integrated the reasoning system onto a Raspberry Pi computer and connected it to the iRobot Create2, powering it from the robot's internal power. We integrated a line sensor and camera and developed the necessary nodes for providing their data to the \gls{BDI} reasoner as perceptions via \gls{ROS}. We provided a translator for the \texttt{create\_autonomy} package, passing sensor data from the robot to the reasoning system, and actions back to the actuators.

\subsection{Future Work} \label{sec:FutureWork}
Our implementation uses line sensing and \gls{QR} to move around the environment and localization. This method was used as a \emph{first iteration} for early development of our prototype system, but it has drawbacks, notably requiring the environment to have line tracks and \gls{QR}. One approach could be to add more docking stations, using them more than just as a charging station but also as guiding beacons (make each station emit a different code to make it distinguishable by the robot), and more generally turning them into full-blown stations for mail drop-off and pick-up. These spots could be placed where wi-fi is accessible so that the robot can receive its missions and notify the recipient that the delivery is ready.
With the prototype working, a revisit to the \gls{ASL} implementation is needed. In the current implementation, the \gls{ASL} code could be refactored to be more idiomatic and less redundant now that we have demonstrated that the underlying components work. Additionally, this includes moving toward the use of the message-passing utilities of Jason. For example, messages from the user should be implemented as messages to the agent, not as perceptions. Another desire is to have multiple robots handling mail delivery together. The robots could work as a team, possibly handing off mail from robot to robot, and managing their battery levels. A user would not summon a specific robot to collect their mail, but would instead request the mail service, which would dispatch a robot to collect mail. From there, the robots could hand off the mail item amongst themselves while working together to deliver all mail that they have within their network. Individual robots may also carry multiple mail items. Lastly, the mobile app can be improved to have maps of segments of the tunnel and estimates for when the mail will be delivered. Furthermore, the implementation of delivery alerts should be completed, and provisions for the safety of mail should be made.

%computer vision methods, for example with OpenCV, and the already mounted camera. This could be used to look for objects, signs and patterns local to a particular area, so that when these are observed, the robot can tell its current location. This could also be used for collision avoidance with pedestrians, other robots, or debris that may be in the tunnels.
    \subsubsection*{Acknowledgement} \label{sec:Acknowledgement}

% Patrick is NSERC funded. This is required in all papers.
% Rule: http://www.nserc-crsng.gc.ca/NSERC-CRSNG/acknowledgement_and_logos-mention_et_logos_eng.asp
We acknowledge the support of the Natural Sciences and Engineering Research Council of Canada (NSERC), [funding reference number 518212].

Cette recherche a \'{e}t\'{e} financ\'{e}e par le Conseil de recherches en sciences naturelles et en g\'{e}nie du Canada (CRSNG), [num\'{e}ro de r\'{e}f\'{e}rence 518212].
    \bibliographystyle{eptcs}
\bibliography{references}
\end{document}